\documentclass[11pt,twoside]{article}
\usepackage{asp2014}
\pdfoutput=1

\aspSuppressVolSlug
\resetcounters

\begin{document}

\title{A New Generation of Cool White Dwarf Atmosphere Models Using Ab Initio Calculations}
\author{Simon~Blouin,$^{1,2}$ Patrick~Dufour,$^1$ and Piotr~M.~Kowalski$^2$
\affil{$^1$Universit\'e de Montr\'eal, Montr\'eal, Qu\'ebec, Canada; \email{sblouin@astro.umontreal.ca}}
\affil{$^2$IEK-6 Institute of Energy and Climate Research, Forschungszentrum J\"ulich, 52425 J\"ulich, Germany}}

\paperauthor{Simon~Blouin}{sblouin@astro.umontreal.ca}{0000-0002-9632-1436}{Universit\'e de Montr\'eal}{D\'epartement de physique}{Montr\'eal}{Qu\'ebec}{H3C 3J7}{Canada}
\paperauthor{Patrick~Dufour}{dufourpa@astro.umontreal.ca}{}{Universit\'e de Montr\'eal}{D\'epartement de physique}{Montr\'eal}{Qu\'ebec}{H3C 3J7}{Canada}
\paperauthor{Piotr~Kowalski}{p.kowalski@fz-juelich.de}{0000-0001-6604-3458}{Forschungszentrum J\"ulich}{IEK-6 Institute of Energy and Climate Research}{J\"ulich}{}{52425}{Germany}

\begin{abstract}
Due to their high photospheric density, cool helium-rich white dwarfs (particularly DZ, DQpec and ultracool) are often poorly described by current atmosphere models. As part of our ongoing efforts to design atmosphere models suitable for all cool white dwarfs, we investigate how the ionization ratio of heavy elements and the H$_2$-He collision-induced absorption (CIA) spectrum are altered under fluid-like densities. For the conditions encountered at the photosphere of cool helium-rich white dwarfs, our ab initio calculations show that the ionization of most metals is inhibited and that the H$_2$-He CIA spectrum is significantly distorted for densities higher than $0.1$\,g/cm$^3$.
\end{abstract}

\section{Introduction}
The photosphere of cool ($T_{\mathrm{eff}} < 6000$\,K) helium-rich white dwarfs reaches fluid-like densities (up to a few~g/cm$^3$) \citep{kowalski200aip}. Under such conditions, collective interactions between particles are no longer negligible, which leads to an alteration of opacities, chemical abundances and the equation of state.

In recent years, many of these non-ideal effects have been investigated. These include corrections to the free-free absorption by He$^-$ and to Rayleigh scattering \citep{iglesias_etal_2002}, high-pressure line profiles for Ca, Mg and Na \citep{allard_alekseev_2014,allard_etal_2016}, high-density UV opacities for hydrogen \citep{kowalski_saumon_2006}, radiative transfer with refraction \citep{kowalski_saumon_2004}, non-ideal dissociation equilibrium of molecular hydrogen \citep{kowalski_2006}, ionization of helium \citep{kowalski_etal_2007}, pressure distortion of $\rm C_2$ Swan bands \citep{kowalski_2010} and infrared absorption of dense helium \citep{kowalski_2014}.

However, even when taking these effects into account, current atmosphere models are unable to reproduce the spectra of several cool DZ, DQpec and ultracool white dwarfs. This suggests that the physics and chemistry included in present atmosphere models is still incomplete. Therefore, to use these stars as cosmochronometers or to access the bulk composition of rocky material accreted on their surface, more fundamental research on the behavior of matter under these conditions is required.

Here we discuss new ab initio calculations that reveal significant deviations from the ideal gas limit for both the ionization of heavy elements and the H$_2$-He collision-induced absorption spectrum.

\section{Non-ideal ionization of heavy elements}
For the ionization of species $Z$ to $Z^+$, minimization of the Helmholtz free energy leads to the condition,
\begin{equation}
\mu_{Z} = \mu_{Z^+} + \mu_{e},
\label{eq:muioniz}
\end{equation}
where $\mu_i$ is the chemical potential of species $i$. In general, $\mu_i$ can be expressed as the sum of an ideal and a non-ideal component,
\begin{equation}
\mu_i = {\mu_i}^{\mathrm{id}} + {\mu_i}^{\mathrm{nid}}.
\end{equation}
The non-ideal component is itself the sum of two terms: an \textit{entropy term}, which reflects the excluded volume of the species and an \textit{interaction term}, which describes the energy of the interaction of the species with surrounding particles. In the low-density limit, ${\mu_i}^{\mathrm{nid}}$ can be neglected and equation \ref{eq:muioniz} leads to the well-known Saha ionization equilibrium equation. However, to properly describe the ionization of metals under the photospheric densities of cool DZ stars, ${\mu_i}^{\mathrm{nid}}$ must be taken into account.

${\mu_e}^{\mathrm{nid}}$ has already been computed by \cite{kowalski_etal_2007} using ab initio methods and agrees well with existing experimental data. To compute ${\mu_Z}^{\mathrm{nid}}$ and ${\mu_{Z^+}}^{\mathrm{nid}}$, we solved the Orstein-Zernike equation following the procedure outlined in \cite{kowalski_etal_2007} using pair potentials computed with the ORCA package \citep{orca} at the CCSD(T) level.

Our results, reported in terms of the ionization ratio, are given in Figure \ref{fig:ioniz}. For every element considered, we find that the ionization fraction deviates from the ideal gas approximation by several orders of magnitude for the conditions encountered at the photosphere of cool DZ stars. Figure \ref{fig:ioniz} shows that at 1\,g/cm$^3$, ionization is inhibited for most heavy elements. This rather unexpected result is primarily due to the high interaction energy of electrons with helium and should be validated by spectral fits or relevant experimental studies.

\articlefigure[width=0.64\textwidth]{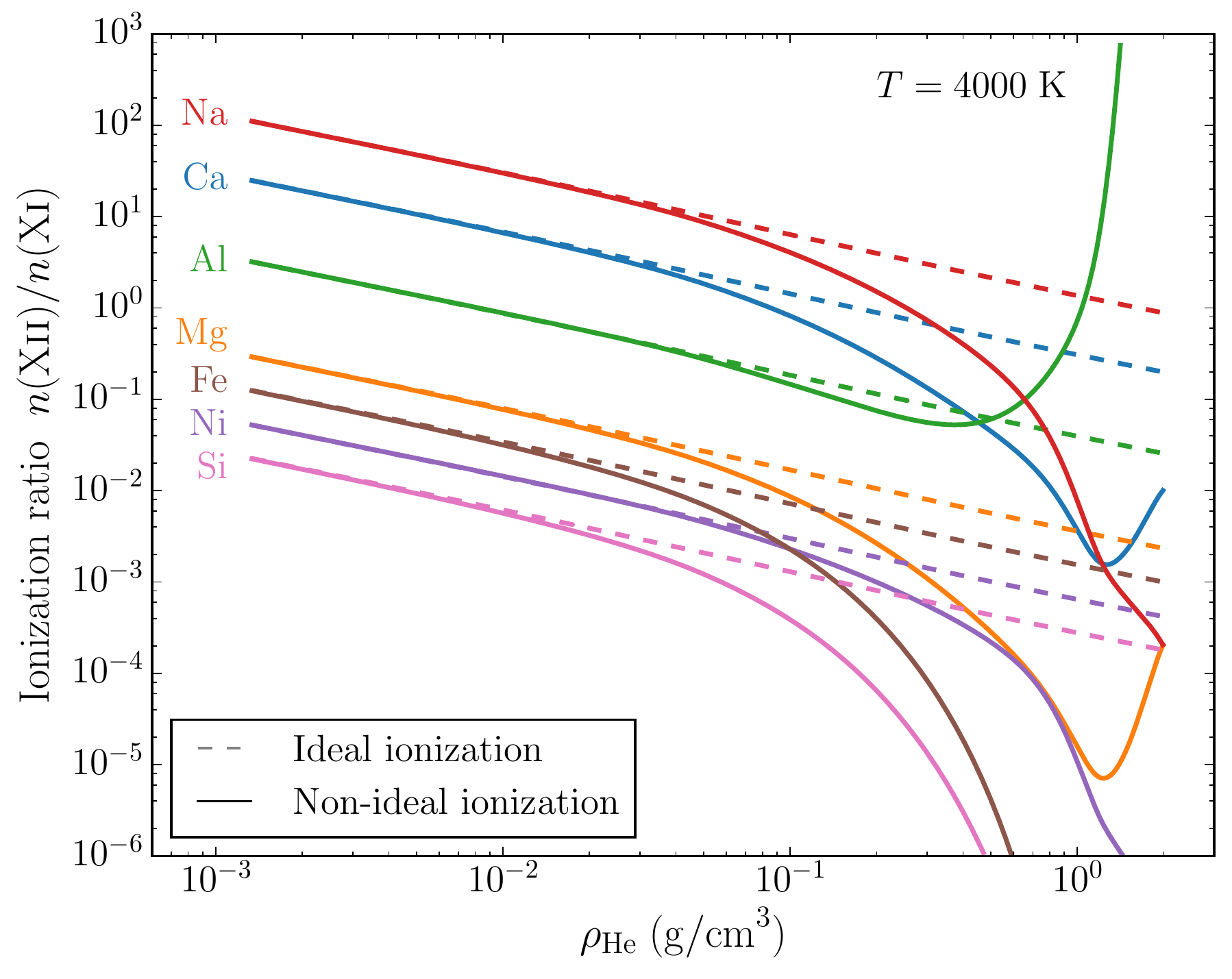}{fig:ioniz}{Ionization ratio of several heavy elements found in DZ stars as a function of helium density. Dashed lines show the prediction of the ideal Saha equation and solid lines show the prediction of our new model (see text).}

\clearpage

\section{H$_\mathbf{2}$-He CIA opacities at high density}
A dipole moment is induced when a collision between particles (in our case, H$_2$ and He) occurs, which results in infrared (IR) absorption known as collision-induced absorption (CIA). To compute CIA opacities, previous studies \citep[e.g.,][]{jorgensen_etal_2000,abel_2012} relied on potential energy and induced dipole surfaces computed in the infinite-dilution limit. At high density (>\,0.1\,g/cm$^3$), this approximation can break down.

To check the effect of many-particle collisions, we performed ab initio molecular dynamics simulations of CIA opacities using the procedure and computational setup of \cite{kowalski_2014}. Numerous simulations were performed for different $(T,\rho)$ conditions. For each simulation, the cubic supercell contained one H$_2$ molecule and either 31 or 63 He atoms and the simulation box length was chosen to obtain the desired density.

For each 0.5\,fs timestep, the total dipole moment $\mathbf{M}$ of the simulation supercell was computed. The IR spectrum was then obtained through the Fourier transform of the dipole moment autocorrelation function \citep{silvestrelli_1997},
\begin{equation}
\alpha (\omega) = \frac{2 \pi \omega^2}{3 c k_B T V} \int_{-\infty}^{\infty} \mathrm{d} t \exp(-i \omega t) \langle \mathbf{M} (t) \cdot  \mathbf{M}(0) \rangle,
\end{equation}
where $\alpha$ is the absorption coefficient, $\omega$ the wavenumber, $c$ the speed of light in the fluid, $k_B$ the Boltzmann constant, $T$ the temperature, 
$V$ the supercell volume and the angle brackets denote the time-correlation function, i.e. $\langle \mathbf{M} (\tau) \cdot \mathbf{M} 
(0) \rangle = \frac{1}{T} \int_0^T \mathbf{M} (t) \cdot \mathbf{M} (t + \tau) \mathrm{d} t$.

\articlefigure[width=0.7\textwidth]{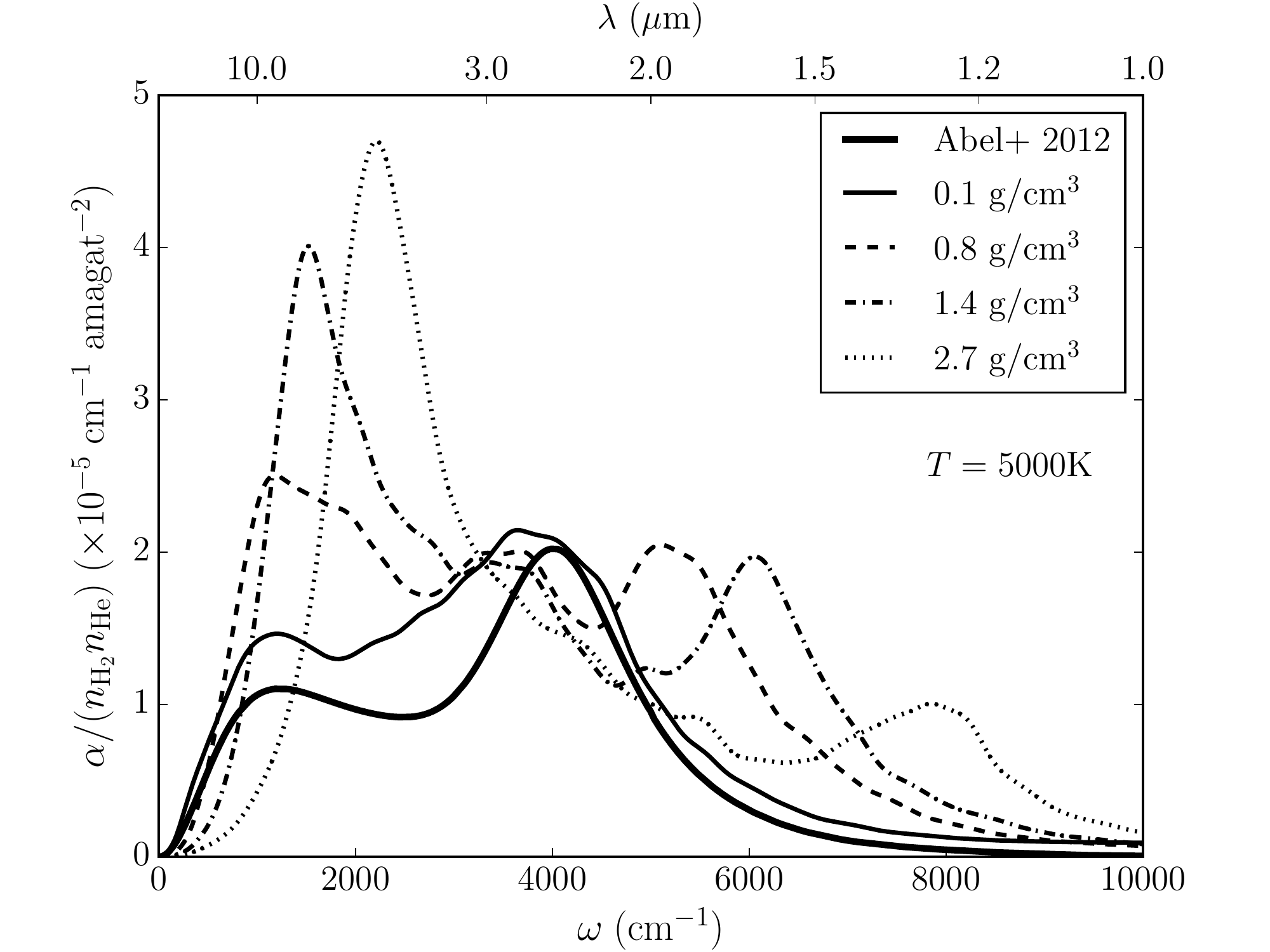}{fig:cia}{The thickest line is the density-normalized H$_2$-He CIA spectrum computed by \cite{abel_2012}, which is valid in the low-density limit. All other lines show the results of our simulations for different helium densities.}

As Figure \ref{fig:cia} shows, the CIA spectrum is significantly distorted for densities encountered at the photosphere of cool helium-rich white dwarfs (>\,0.1\,g/cm$^3$). This affects the near-IR and IR spectra of cool white dwarfs, as discussed by Kowalski et al. in these proceedings. We found that the length of the H$_2$ bond is reduced by $\approx$\,15\,\% when density is raised from 0.1 to 2\,g/cm$^3$. This compression of the H$_2$ molecule shifts its rovibronic levels, which might explain the CIA spectrum distortion. In addition, under high-density conditions, triple collisions become important and increase the absorption coefficient \citep{lenzuni_saumon_1992}. From Figure \ref{fig:cia}, it is also apparent that the fundamental band is split at high density. This has already been experimentally observed \citep{hare_welsh_1958} and it is explained by the interference resulting from correlations between dipole moments induced between successive collisions \citep{kranendonk_1968}.

\section{Conclusion}
Using state-of-the-art ab initio calculations, we find yet unaccounted high-density effects that alter the ionization equilibrium of heavy elements and the shape of the H$_2$-He CIA spectrum in the atmosphere of cool helium-rich white dwarfs. These non-ideal effects will modify the synthetic spectra computed by atmosphere models and, therefore, effective temperature and chemical abundances inferred from observations will also be affected. Many more non-ideal effects still have to be taken into account (e.g., additional metal line profiles at high density, non-ideal dissociation of C$_2$, etc.). Once all the most important non-ideal effects are included in our models, they will be used to properly analyse cool DZ, DQpec and ultracool white dwarfs.

\acknowledgements S.B. acknowledges support from NSERC (Canada), FRQNT (Qu\'ebec) and DAAD (Germany).

\end{document}